\documentclass[pre,showpacs]{revtex4}
\usepackage{amssymb}
\usepackage[dvips]{graphicx}
\usepackage{amsmath}
\usepackage{graphicx}
\usepackage{makeidx}
\usepackage{subfigure}
\usepackage{amsfonts}
\usepackage{times}
\usepackage{epsfig}

\begin{document}

\title{Dynamics of bright solitons and soliton arrays 
in the nonlinear Schr{\"{o}}dinger equation with
a combination of random and harmonic potentials}
\author{Qian-Yong Chen and Panayotis G. Kevrekidis}
\affiliation{Department of Mathematics and Statistics, University of Massachusetts,
Amherst MA 01003-4515, USA}
\author{Boris A. Malomed}
\affiliation{Department of Physical Electronics, School of Electrical Engineering,
Faculty of Engineering, Tel Aviv University, Tel Aviv 69978, Israel}
\affiliation{ICFO-Institut de Ciencies Fotoniques, Mediterranean Technology Park,
Castelldefels 08860, Spain\footnote{%
Sabbatical address}}

\begin{abstract}
We report results of systematic simulations of the dynamics of
solitons in the framework of the one-dimensional nonlinear
Schr\"{o}dinger equation (NLSE), which includes the
harmonic-oscillator (HO) potential and a random potential. The
equation models experimentally relevant spatially disordered
settings in Bose-Einstein condensates (BECs) and nonlinear optics.
First, the generation of soliton arrays from a broad initial
quasi-uniform state by the modulational instability (MI) is
considered, following a sudden switch of the nonlinearity from
repulsive to attractive. Then, we study oscillations of a single
soliton in this setting, which models a recently conducted
experiment in BEC. Basic characteristics of the MI-generated array,
such as the number of solitons and their mobility, are reported as
functions of the strength and correlation length of the disorder,
and of the total norm. For the single oscillating soliton, its survival
rate is found. Main features of these dependences are explained
qualitatively.
\end{abstract}

\maketitle

\section{Introduction}

The interplay of the disorder and nonlinearity is a topic which has been
attracting a great deal of interest for a long time, see reviews \cite%
{review}-\cite{review3}. In particular, much work has been done on the
analysis of dynamics of solitons in disordered external potentials, chiefly
in one-dimensional settings, see, e.g., Refs. \cite{soliton1}-\cite{review4}
and references therein. These studies find an important application to the
description of Bose-Einstein condensates (BECs) trapped in random
potentials. The latter topic was addressed in many theoretical \cite{BEC0}-%
\cite{BEC6} and experimental \cite{Randy1}-\cite{Randy3} works.

A ubiquitous theoretical model used in this field is the nonlinear Schr\"{o}%
dinger equation, NLSE (alias the Gross-Pitaevskii equation, in terms of BEC
\cite{GPE}) for the wave function $u(x,t)$, which includes a regular trapping
potential and a term accounting for a disordered potential. The normalized
form of this equation, with scaled time $t$ and space 
coordinate $x$, is%
\begin{equation}
iu_{t}+\frac{1}{2}u_{xx}+g|u|^{2}u-\frac{1}{2}{\Omega }^{2}x^{2}u-V_{\mathrm{%
dis}}(x)u=0,  \label{NLS}
\end{equation}%
where $g=+1$ and $-1$ correspond to the self-attractive and self-repulsive
nonlinearities, both signs being possible in BEC, $\Omega$
characterizes 
the
strength of the harmonic-oscillator (HO) trapping potential, and $V_{\mathrm{%
dis}}(x)$ is a random potential \ representing the spatial disorder. In
particular, this model describes dipole oscillations of the trapped
condensate, experimentally studied in recent work \cite{Randy3}, where
strong effective dissipation induced by the disordered potential was
discovered. On the other hand, Eq. (\ref{NLS}) with $t$ replaced by the
propagation distance
coordinate $z$ is a model of a nonlinear optical waveguide with
a random perturbation of the local refractive index, which is represented by
the disordered potential \cite{review0}, hence the results reported below
apply to spatial solitons in the nonlinear waveguide as well.

Our objective is to systematically investigate fundamental properties of
solitons in the framework of the model based on Eq. (\ref{NLS}). These
include the formation and motion of soliton trains due to the
modulational instability (MI) of the broadly distributed condensate, and, as
suggested by the experiment reported in Ref. \cite{Randy3}, oscillations of
a single soliton which is originally placed at a distance from the minimum
of the trapping potential, $x=0$. The results for these two types of the
dynamical behavior, based on systematic simulations of Eq. (\ref{NLS}) and
averaging over many realizations of the random setting, as well as on a
qualitative analysis of the observed effects, are reported, respectively, in
Sections II and III, and the paper is concluded by Section IV.

\section{Formation of soliton arrays by the modulational instability in the
spatially disordered environment}

\label{sec:1D}

The subject of this Section is the effect of the spatial disorder on the
formation of soliton chains from an initial quasi-uniform state, and the
subsequent evolution of the chains. The numerical simulations were subject
to the following specifications. The spatial domain was taken as $-60\pi
<x<+60\pi $, and the trapping strength is $\Omega ^{2}=25\times 10^{-4}$,
hence the respective HO length
%, 
%$x_{0}=\sqrt{2/\Omega }\approx 6.3$, 
is much
smaller than the size of the integration domain, which makes it possible to
consider multi-soliton trains, and persistent motion of the solitons. The
split-step Fourier method with a spatial grid composed of $1024$ points
and time step $\Delta t=0.001$ was employed for the spatial and temporal
discretization of Eq. (\ref{NLS}). The total integration time is $T=80$.
 A fourth-order optimized implementation of the
splitting~was used, as in Refs. \cite{Blanes02, Montesinos05}. Further, the
disorder was represented by a spatially correlated random function, $V_{%
\mathrm{dis}}(x)=V_{d}f(x,\theta )$, where $V_{d}$ is the strength of the
disorder, and the marginal distribution of $f(x,\theta )$ has an exponential
form, its covariance function being $C(x_{1},x_{2})=\exp
(-2(x_{1}-x_{2})^{2}/V_{z}^{2})$, in which $V_{z}$ is the correlation
length. 
Other type of random functions as in~\cite{Yaglom}  can be considered in a similar way.

The initial quasi-uniform distribution of the condensate is taken as the
ground-state solution of Eq. (\ref{NLS}) with the \emph{self-repulsive}
nonlinearity, i.e., with $g=-1$, in the form of $u=\exp \left( -i\mu
t\right) U(x)$, where $\mu $ is the real chemical potential, and function $%
U(x)$ was found as a stationary solution of the associated nonlinear
diffusion equation (the imaginary-time version \cite{imaginary} of Eq. (\ref%
{NLS})):
\begin{equation}
u_{t}=\frac{1}{2}u_{xx}-|u|^{2}u-\frac{1}{2}\Omega ^{2}x^{2}u-V_{\mathrm{dis}%
}(x)u+\mu u.  \label{diff}
\end{equation}%
The same split-step method was used to solve Eq. (\ref{diff}). Then, the
evolution of the MI (modulational instability) of this state was simulated
in real time, following a sudden switch of the self-repulsion into
self-attraction, i.e., replacement of $g=-1$ by $g=+1$ in Eq. (\ref{NLS}),
which exactly corresponds to the reversal of the sign of the nonlinearity by
means of the Feshbach resonance in the well-known BEC experiment 
of \cite{Randy}.

%Even without the disorder, stable localized soliton exist when the sign of the nonlinearlity
%is switched from $-1$ to $1$. This will be our benchmark case. One such result is shown in Fig.~\ref{f:benchmark}.
%
% \begin{figure}
%      \centerline{
%               \psfig{file = Nx4096_dt001.eps, width=3in,height=3in}
%           }
%
%      \caption{Reference solution with absence of disorder}
%      \label{f:benchmark}
%  \end{figure}

The MI splits the initial quasi-uniform state into a chain of
solitons, which, generally speaking, is an obvious outcome of the
evolution induced by the reversal of the nonlinearity sign. However,
a new aspect of this outcome, on which we focus here, is the effect
of the disordered environment on the resulting solitary wave chain. 
We have collected results which
demonstrate the dependence of the characteristics of the emerging
soliton chain on three control parameters: strength $V_{d}$ and
correlation length $V_{z}$ of the random potential, and the total
norm of the initial state, $N=\int_{-\infty }^{+\infty
}|u(x)|^{2}dx$. The characteristics whose variation was monitored
are (i) the number of solitons; (ii) the largest displacement of
solitons, i.e., the largest distance that peaks of individual
solitons can travel in the course of the evolution; Basically the displacement for each soliton
is computed as the distance between its leftmost and rightmost positions. Then the largest displacement
is simply the largest value among all the solitons.
(iii) the normalized average kinetic energy per soliton, which is
defined as
\begin{equation}
\bar{K}=\left( \sum M_{j}\right) ^{-1}\left( \sum \frac{1}{2}%
M_{j}v_{j}^{2}\right) ,  \label{K}
\end{equation}%
where the summation is performed on the full set of solitons in the emerging
configuration, $v_{j}$ is the velocity of the $j$-th soliton, and $%
M_{j}=\int |u(x)|^{2}dx$ is its effective mass, with the integration
performed over a vicinity of the solitary wave's
peak where the local amplitude of
the field exceeds half of its peak value. 

\begin{figure}[tbp]
\centerline{
                \psfig{file = 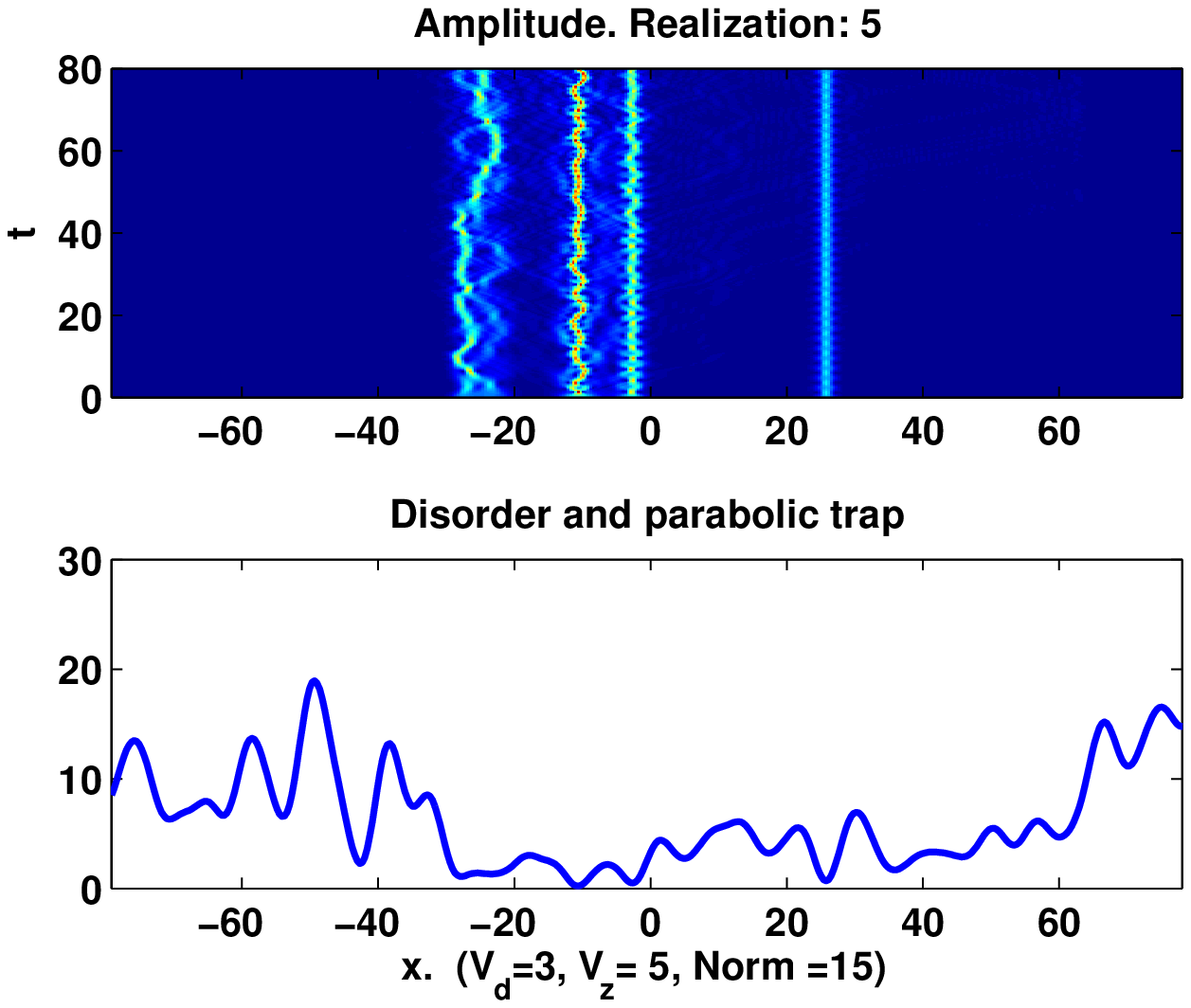, width=2.5in,height=2.5in}
                \hspace{0.1in}
                \psfig{file = 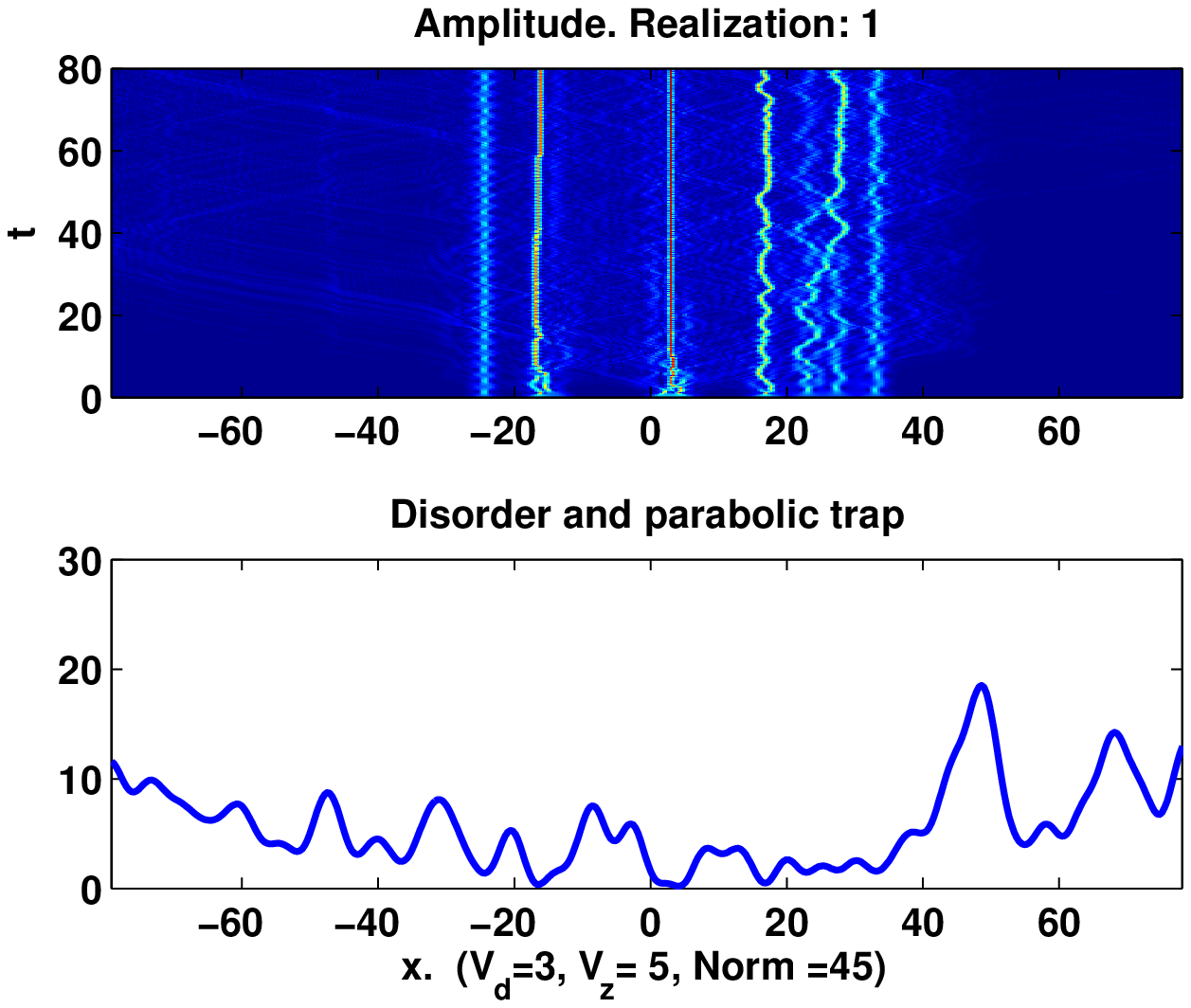, width=2.5in,height=2.5in}}
\caption{(Color online) Two realizations of the random multi-soliton
configuration generated by the development of the modulational instability
for $V_{z}=5$. Here and in the next figure, top plots depict the
spatiotemporal evolution of $|u(x,t)|$, while the bottom ones display the
corresponding total potential---the sum of the HO trap and the random part.}
\label{f:VZ5}
\end{figure}

\begin{figure}[tbp]
\centerline{
                \psfig{file = 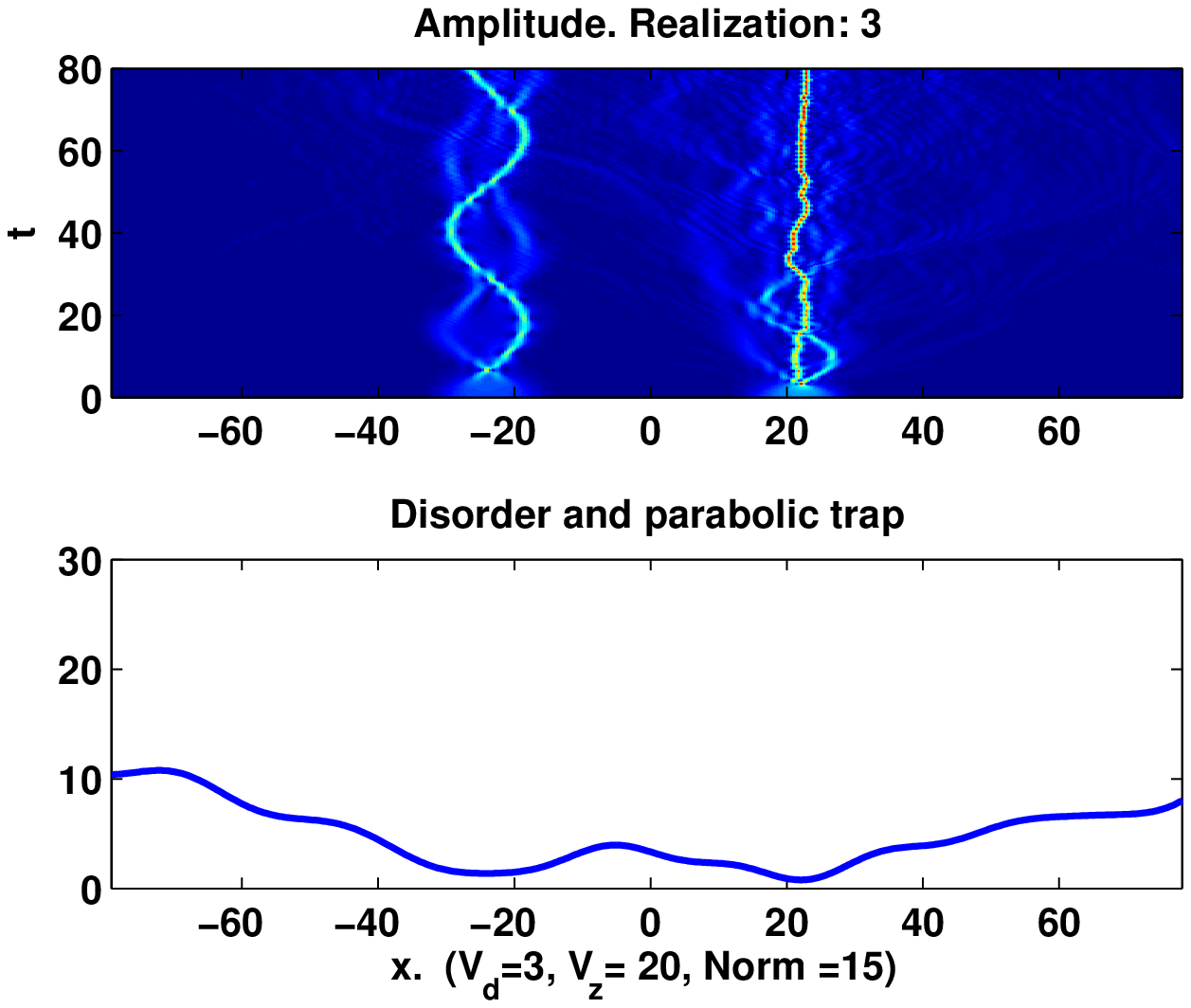, width=2.5in,height=2.5in}
                \hspace{0.1in}
                \psfig{file = 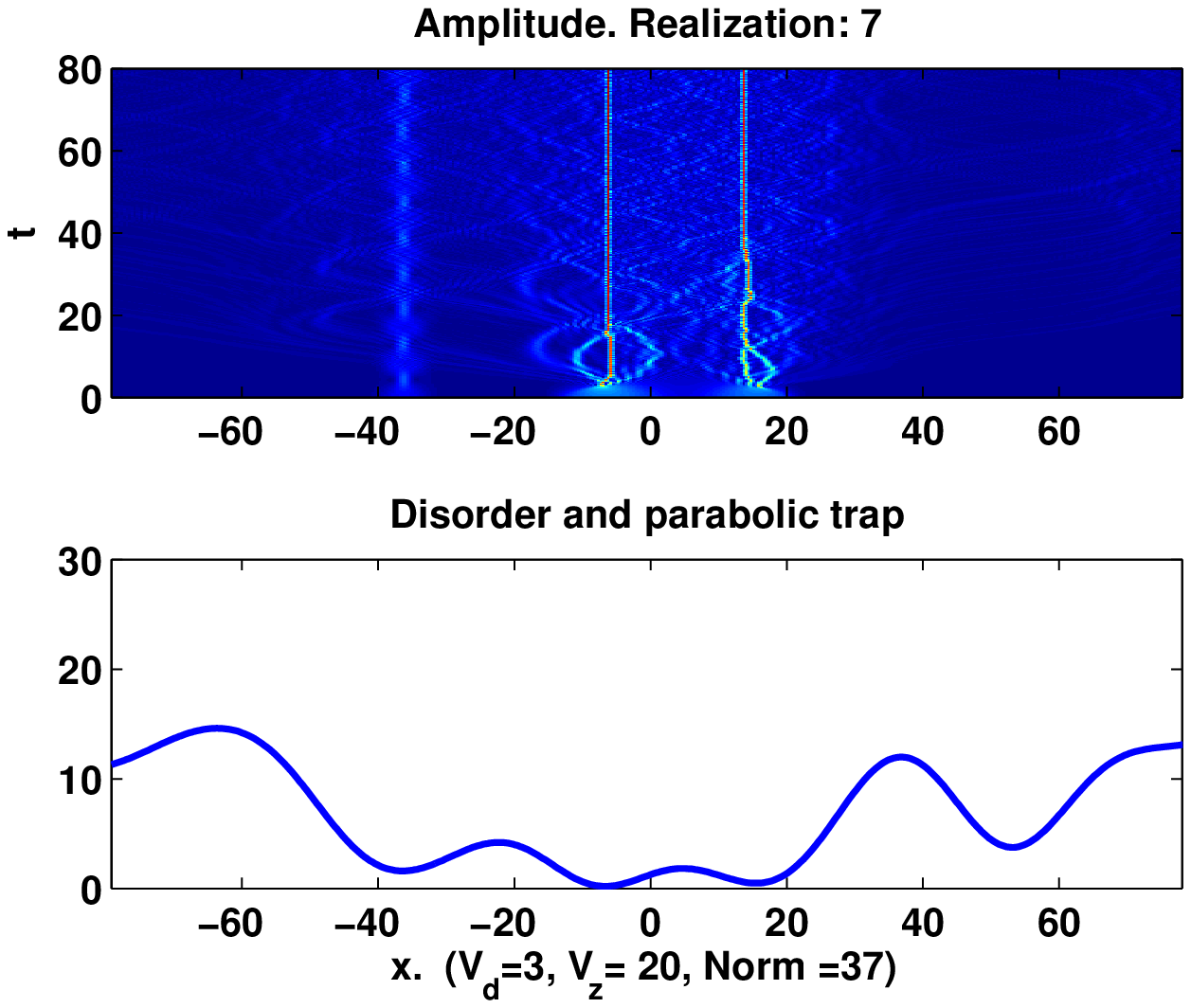, width=2.5in,height=2.5in}}
\caption{(Color online) Two realizations of the random multi-soliton
patterns for $V_{z}=20$}
\label{f:VZ20}
\end{figure}

The results presented below were produced by averaging over 100
different random realizations. Note that this 
leads to a non-integer number 
as the number of solitons for most cases. Typical examples of the realizations
are displayed in Figs. \ref{f:VZ5} and \ref{f:VZ20}, for small and
large correlations lengths, $V_{z}=5$ and $V_{z}=25$, respectively.
Individual solitons can be easily identified in the plots.

\begin{figure}[tbp]
\centerline{
                \psfig{file = 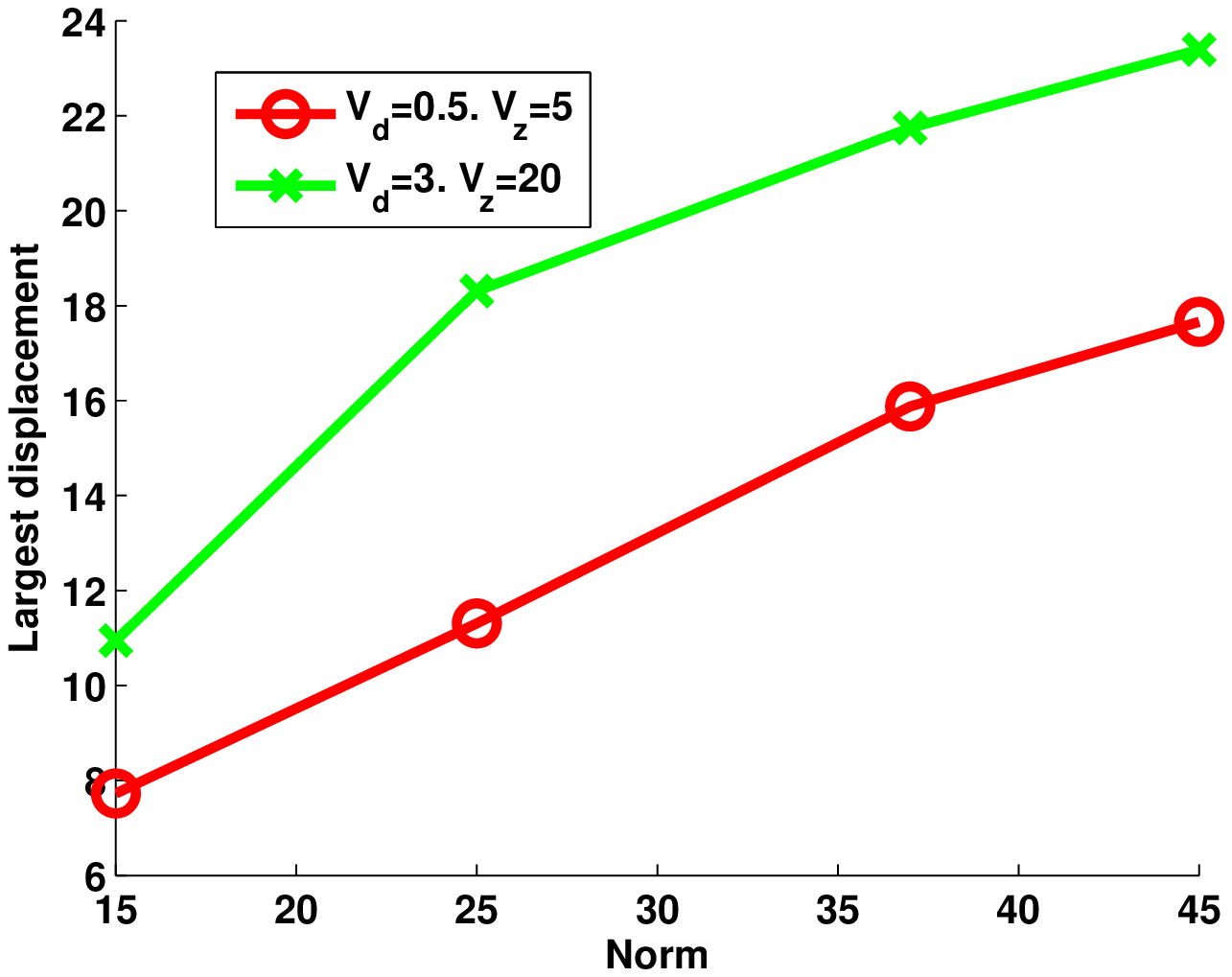, width=2.5in,height=2.5in}
                \hspace{0.1in}
                \psfig{file = 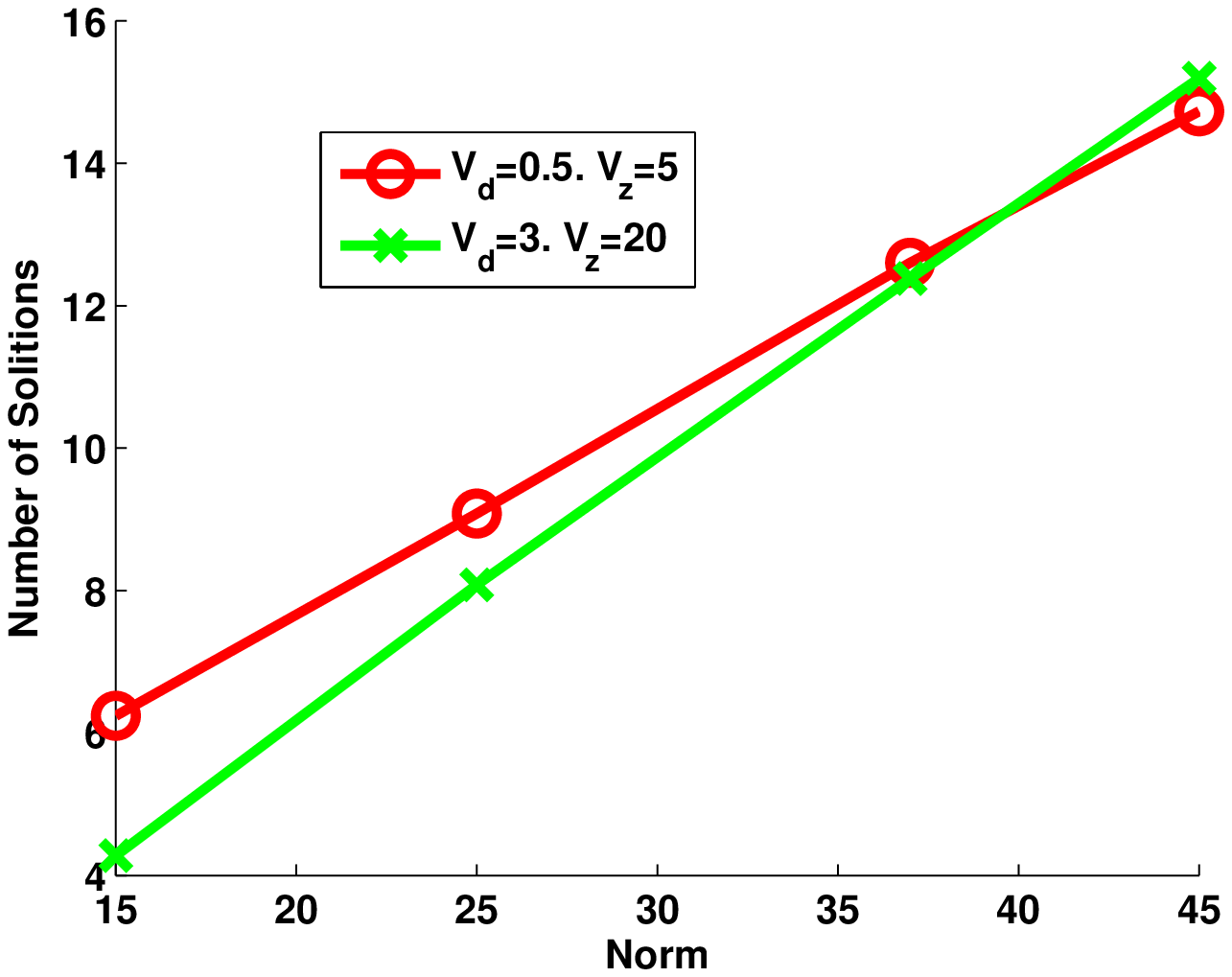, width=2.5in,height=2.5in}
            }
\caption{(Color online) Results obtained for fixed strength and correlation
length of the disorder, while the total norm varies. Here and in the two
following figures, the left and right panels display, respectively, the
largest displacement of solitons, and the number of solitons in the final
configuration produced by the simulations.}
\label{f:varyNorm}
\end{figure}

\begin{figure}[tbp]
\centerline{
                \psfig{file = 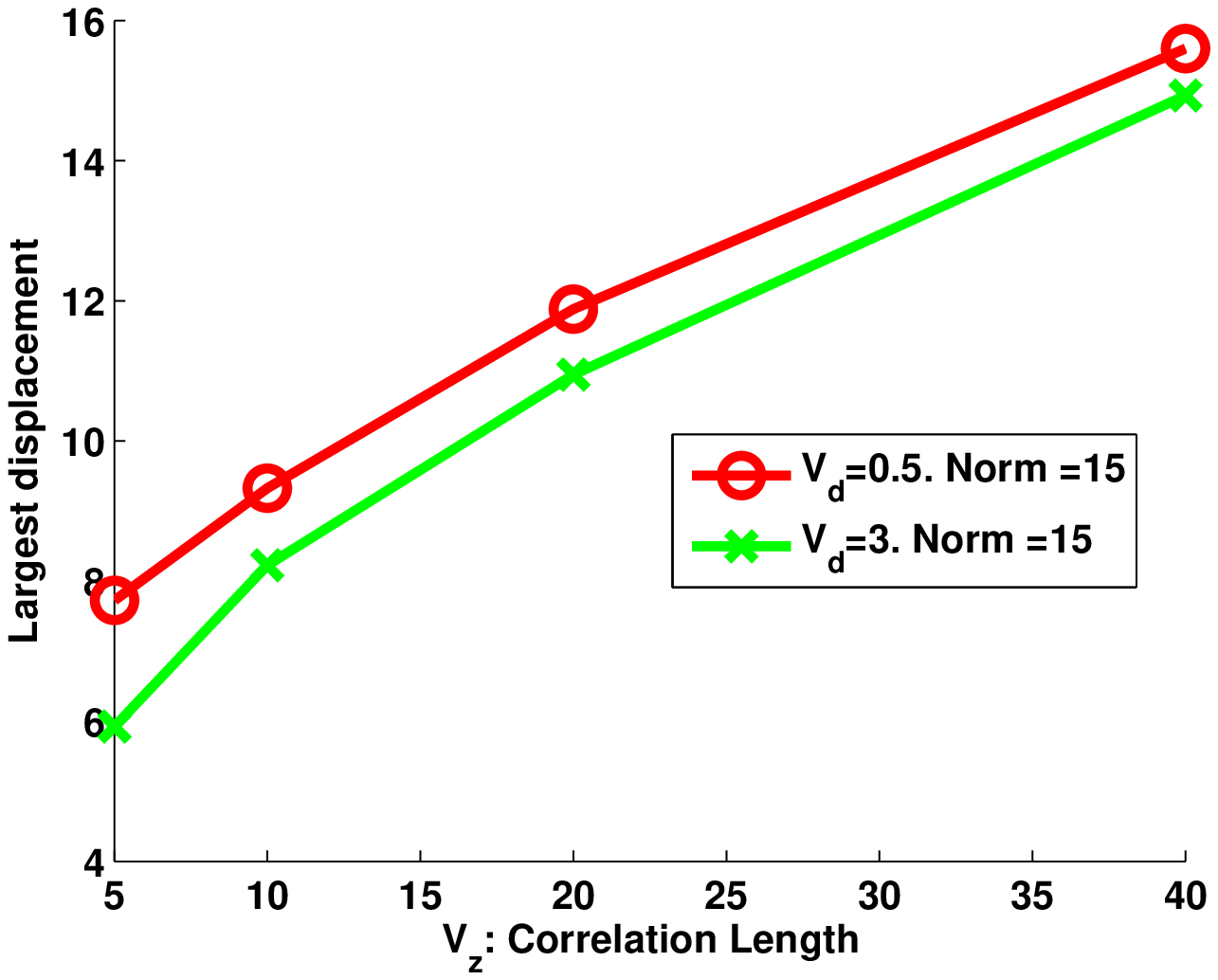, width=2.5in,height=2.5in}
                \hspace{0.1in}
                \psfig{file = 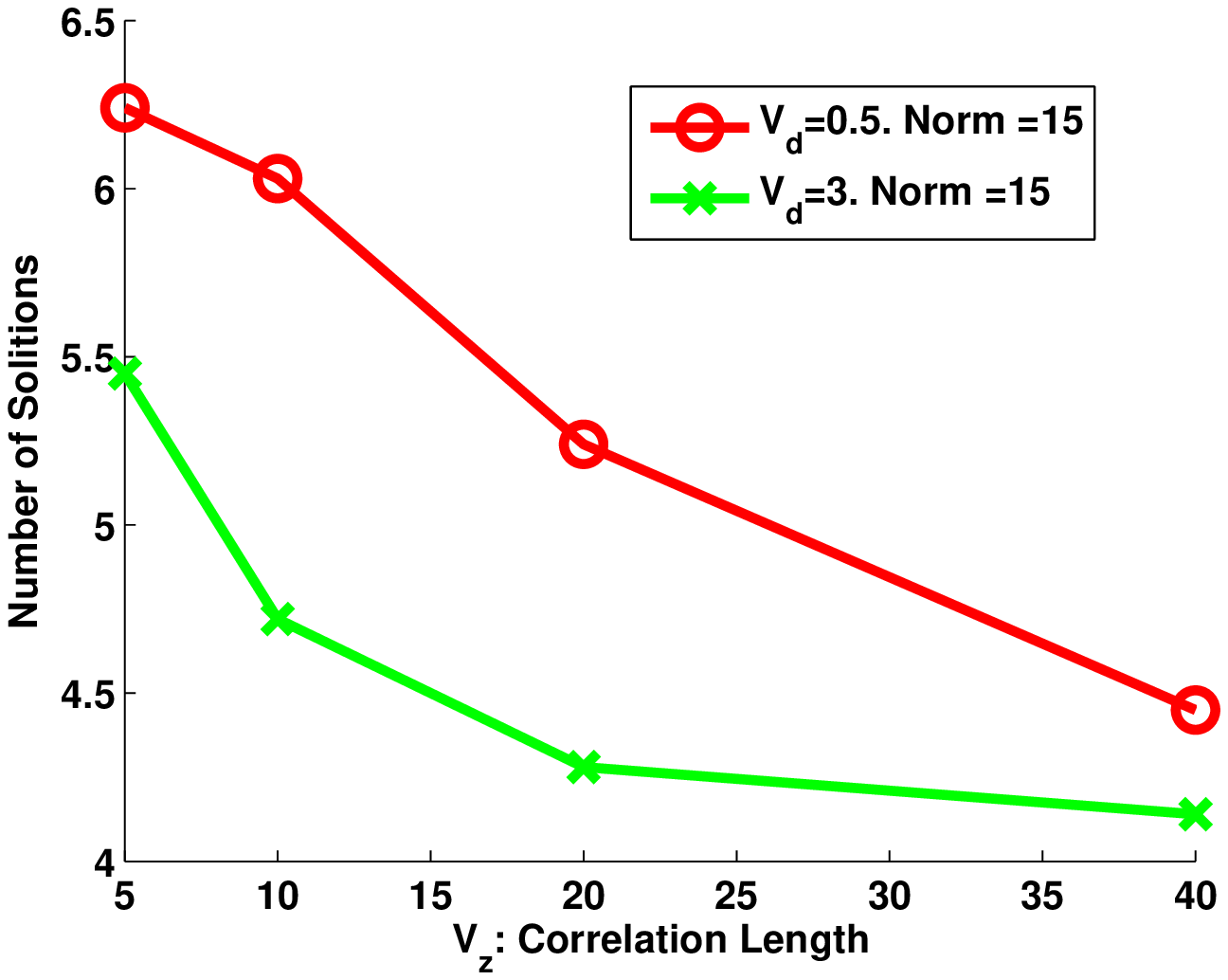, width=2.5in,height=2.5in}
            }
\caption{(Color online) The results for fixed total norm and strength of the
disorder.}
\label{f:varyLength}
\end{figure}

\begin{figure}[tbp]
\centerline{
                \psfig{file = 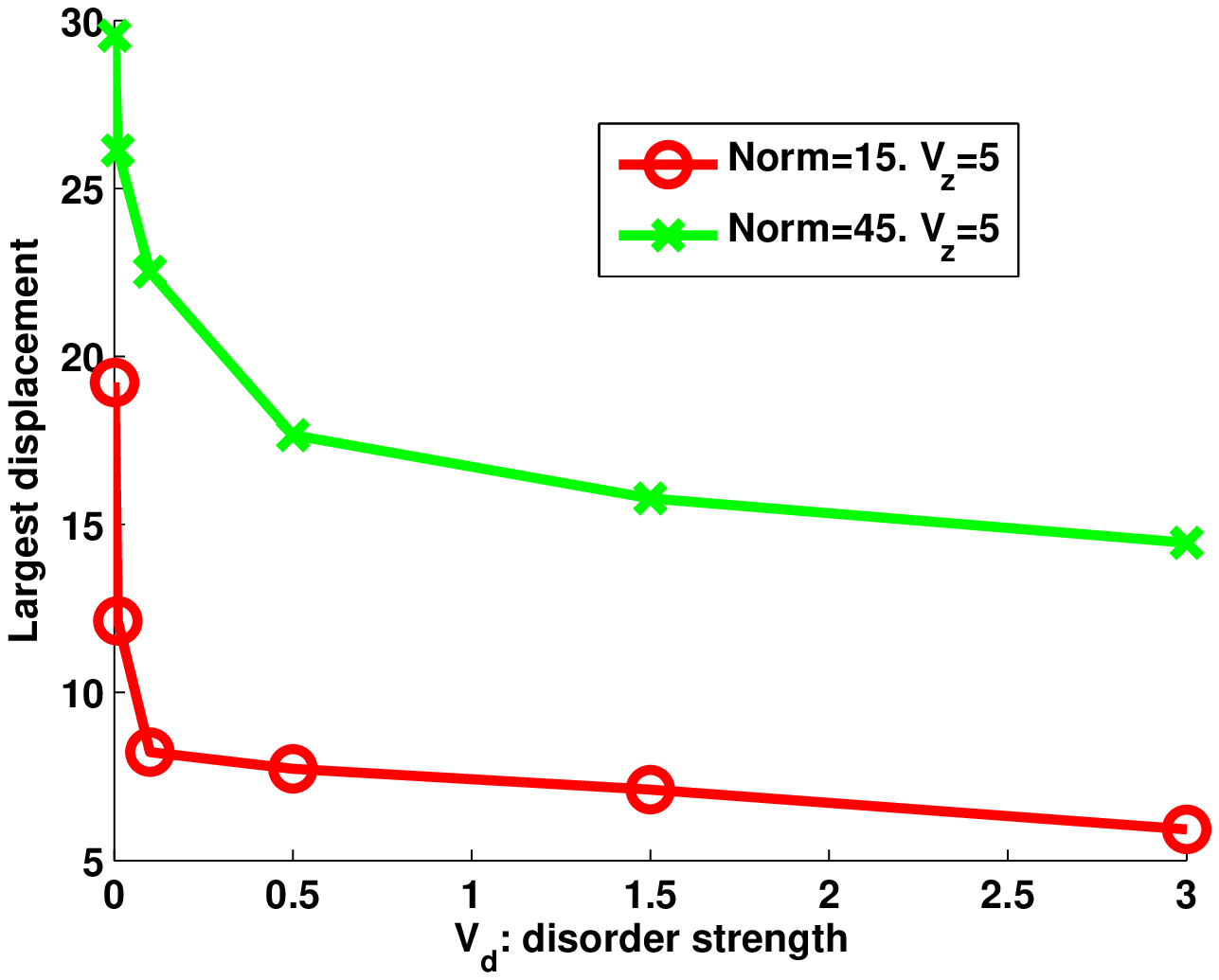, width=2.5in,height=2.5in}
                \hspace{0.1in}
                \psfig{file = 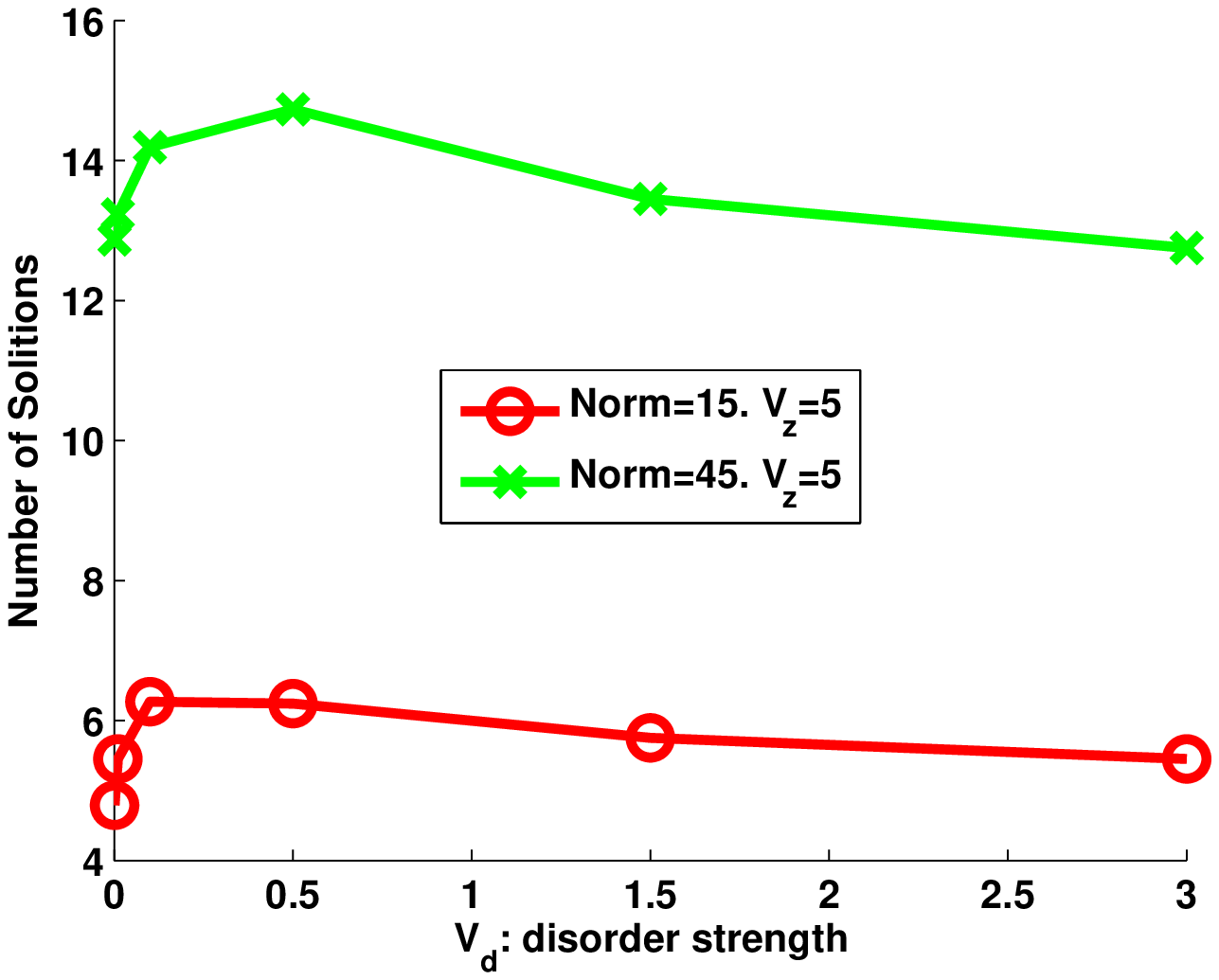, width=2.5in,height=2.5in}
            }
\caption{(Color online) The results for fixed total norm and correlation
length of the disorder.}
\label{f:varyStrength}
\end{figure}

The results for the number of solitons in emerging pattern and their largest
displacement are summarized in Figs.~\ref{f:varyNorm} - \ref{f:varyStrength}%
. In each panel, we fix two of the above-mentioned control parameters ($%
V_{d},V_{z},N$) and vary the third one.
%To see the intrinsic trend of the interested quantities,
%we plot the results for four different configurations of the first two parameters.
For example, the norm is varied in Fig.~\ref{f:varyNorm}, each curve
corresponding to a particular set of values of $V_{d}$ and $V_{z}$.

\begin{figure}[tbp]
\centerline{
                \psfig{file = 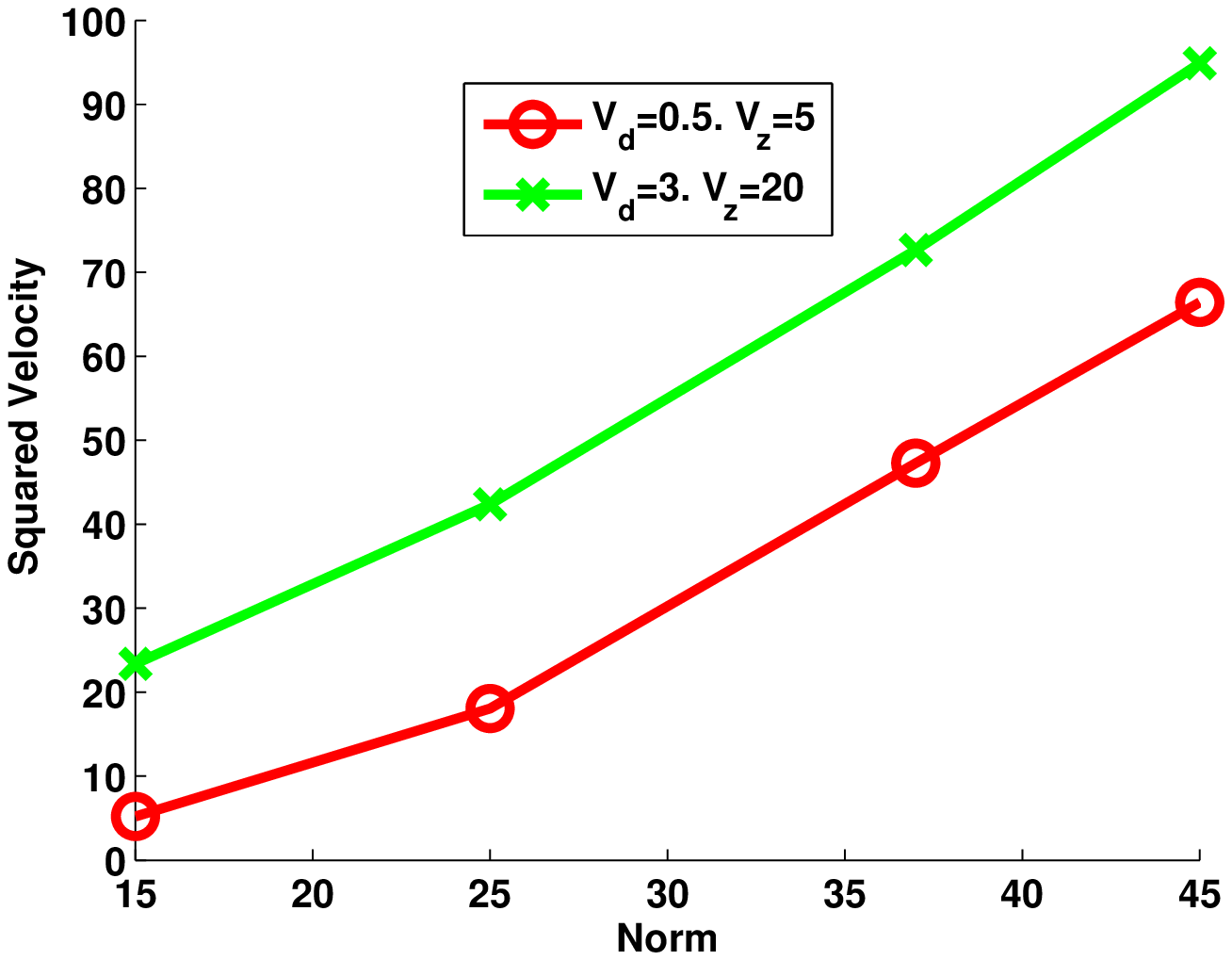, width=2.5in,height=2.5in}
            }
\caption{(Color online) The variation of the mean kinetic energy with
increase of the total norm. }
\label{f:energy}
\end{figure}

The following conclusions can be made from Figs. \ref{f:varyNorm} - \ref%
{f:varyStrength}.

\begin{enumerate}
\item According to Fig.~\ref{f:varyNorm}, the number of solitons in the
chain increases linearly with the total norm, so that the mean norm per
soliton, $N_{\mathrm{sol}}$, is approximately constant. This is a direct
effect of trapping the wave field by the random potential: in free space,
the entire condensate tends to coalesce into a single soliton, which
corresponds a minimum of the system's Hamiltonian \cite{Zakharov}. However,
the sufficiently strong disorder pins portions of the fragmented condensate,
forcing them to self-trap into separated solitons. A fundamental threshold
condition for the self-trapping of an initial state into an NLS soliton is
known in the form of the condition imposed on the \textit{area} of the
initial configuration \cite{Zakharov}:
\begin{equation}
S\equiv \int_{-\infty }^{+\infty }|u_{0}(x)|dx>S_{0}\equiv \ln \left( 2+\sqrt{%
3}\right) \approx 1.32.  \label{A}
\end{equation}%
For fixed parameters of the disorder, i.e., fixed average width $\bar{L}$ of
local potential wells trapping fragments of the condensate, the
above-mentioned constancy of the norm-per-soliton, $N_{\mathrm{sol}}$,
implies a constant average amplitude, $\bar{A}\sim \sqrt{N_{\mathrm{sol}}/%
\bar{L}}$, hence the average area per soliton is constant too, $\bar{S}\sim
\bar{A}\bar{L}\sim \sqrt{N_{\mathrm{sol}} \bar{L}}$, which is consistent
with condition (\ref{A}), that also implies an approximately constant area
per soliton. Thus, condition (\ref{A}) may explain the results observed in
Fig.~\ref{f:varyNorm}.

\item Figure~\ref{f:varyNorm} demonstrates the increase of both the largest
distance traveled by the solitons, and their mean kinetic energy, with the
increase of the total norm. This feature may be explained by the fact that
an individual soliton, moving through the disordered potential, is strongly
braked due to the emission of radiation \cite{review}. The effectively
dissipative character of the motion of coherent wavepackets in this setting was
recently demonstrated in the experiment of \cite{Randy3} (see also the
following Section). However, if the system is filled by the trapped
condensate, the moving soliton actually interacts with trapped segments of
the condensate, i.e., with an effective \textit{pseudopotential }\cite%
{review4}, which is essentially smoother than the ``bare" random
potential. This effect leads to the suppression of the radiation
losses, allowing the solitons to be more mobile.

\item Figure~\ref{f:varyLength} shows that the soliton number decreases,
while the largest displacement increases, as the correlation length of the
disorder, $V_{z}$, increases. This conclusion is consistent with the
conclusions presented in the previous item, as the increase of $V_{z}$
implies the transition to a less disordered potential.

\item Figure~\ref{f:varyStrength} clearly shows that both the soliton number
and largest displacement change very rapidly when $V_{d}$ increases from
zero to small finite values, i.e., the disorder starts to kick in when its
strength is quite small. The observed jump of the soliton number to larger
values, and the simultaneous drop of the free-path length are consistent
with the above argument which states that a deeper random potential splits
the condensate into a large number of solitons, and impedes their free
motion.
\end{enumerate}

%The dark soliton case is also considered. More precisely, we change it to
%the defocusing case ($g=-1$) at time $t=20$ during the simulation. The
%space-time solutions for two configurations are drawn in Fig.~\ref%
%{f:defocusing1} and~\ref{f:defocusing2}, side-by-side with the solutions for
%the focusing case.

%\begin{figure}[tbp]
%\centerline{
%$                \psfig{file = VD05_VZ5_Norm15_MC1.eps,
%width=2.5in,height=2.5in}\hspace{0.1in}
%\psfig{file = VD05_VZ5_Norm15_MC1_Defocusing.eps, width=2.5in,height=2.5in}
%            }
%\caption{Initial norm: $15$; $V_{z}=5$; $V_{d}=0.5$. Left: Focusing; Right:
%Defocusing.}
%\label{f:defocusing1}
%\end{figure}
%
%\begin{figure}[tbp]
%\centerline{
%                \psfig{file = VD3_VZ5_Norm45_MC2.eps, width=2.5in,height=2.5in}
%                \hspace{0.1in}
%                \psfig{file = VD3_VZ5_Norm45_MC2_Defocusing.eps, width=2.5in,height=2.5in}
%            }
%\caption{Initial norm: $45$; $V_{z}=5$; $V_{d}=3$. Left: Focusing; Right:
%Defocusing.} \label{f:defocusing2}
%\end{figure}

\section{Oscillations of the soliton in the combined potential}

Our next objective is to simulate the oscillatory motion of a single soliton in
the combined HO an random potentials, which parallels the recently
reported experiment performed in the condensate of $^{7}$Li
\cite{Randy3}. While the latter experiment focused on the repulsive
interaction case of the (dissipative in the presence of disorder)
dipolar motion of a full condensate, it is straightforward to 
envision such dynamics for an  attractive interaction condensate,
namely a localized solitary wave.
For this purpose, simulations of Eq. (\ref{NLS}) were run with parameters
selected as the rescaled version of those dealt with in the experiment,
where the total number of atoms was $\simeq 10000$, the scattering length $%
a_{s}$ is taken as three times the Bohr radius of $^{7}$Li, the
transverse trapping frequency is $\omega _{r}=2\pi \times 260$ Hz, and the
longitudinal one is $\omega _{z}=2\pi \times 5.5$ Hz.
% NAtom = 10000                   % # of atoms
% a_s = 3*Bohr                    % scattering length
% omega_z = 2*pi*5.5;             % axial frequency, Hz
% omega_r = 2*pi*260;             % perpendicular frequency, Hz
%
%VD = 280/omega_r               % disorder strength
%sigma_D = 23.57 micrometer  % disorder correlation length
%FWHM =  9.4 micrometer         % for IC
%A = 0.6mm                        % initial center of soliton
 The initial conditions were taken as
$u_{0}(x)=\sqrt{2a_{\mathrm{IC}}}{\rm
sech}(\sqrt{2a_{\mathrm{IC}}}(x-x_{0}))$, where $x_{0}$ is the
initial shift of the soliton from the bottom of the HO potential,
and $a_{\mathrm{IC}}$ is determined by the total number of atoms.
With the above physical values, 
$\Omega ^{2}=4.4749 \times 10^{-4}$ and $a_{\mathrm{IC}}=0.2269$ were used in the simulations.
The split-step Fourier method was employed in this case too, with a
sufficiently small spacing in order to properly resolve the size of
the
soliton.
%---roughly, with 10 points covering the FWHM\
%(full-width-at-half-maximum) size of the soliton. 
In the course of
the simulations, the disorder in Eq. (\ref{NLS}) was turned on after
one and a quarter of the period of the oscillations of the soliton
in the HO potential, i.e., when the soliton's center was passing the
origin,
$x=0$.

\begin{figure}[tbp]
\centerline{
    \psfig{file =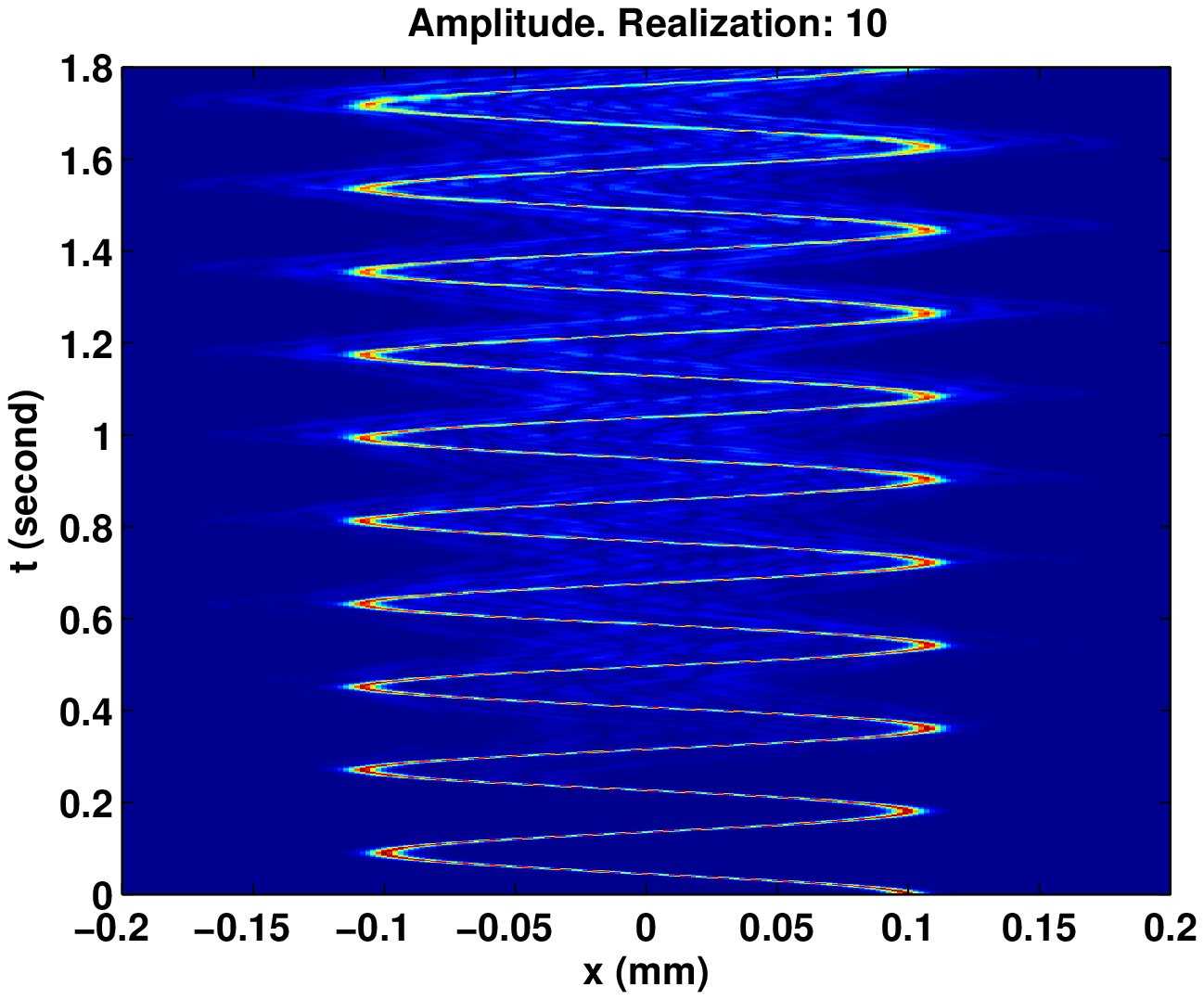,  width=2.5in,height=2.5in}
\hspace{0.1in} \psfig{file =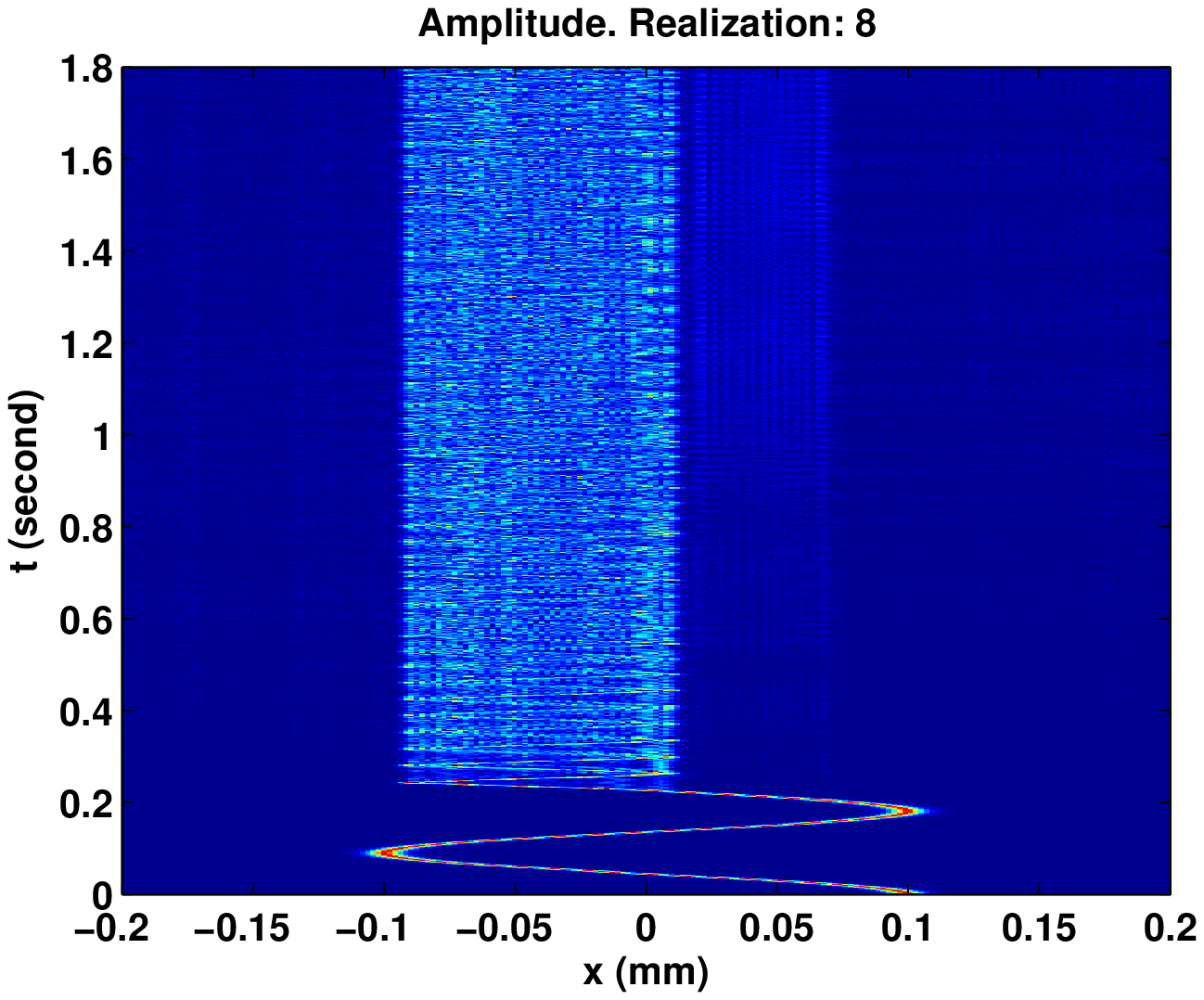,
width=2.5in,height=2.5in}
      }
\caption{(Color online) The left and right panels display examples of solitons
which, respectively, survive and get destroyed in the course of the
oscillations, for an initial shift of the soliton to $x_{0}=0.1~$mm, its
FWHM width $9.2142~\mathrm{\protect\mu }$m, and a
disorder correlation length of $%
V_{z}=9.4264~\mathrm{\protect\mu }$m. The disorder strength corresponding to
the left and right panels is, respectively, $V_{d}=h \cdot 5$(Hz) and $h \cdot 180$%
(Hz), where $h$ is the Planck constant.}
\label{f:VD5_180}
\end{figure}

For given initial shift $x_{0}$, we mainly varied two parameters (as
before), the correlation length $V_{z}$ and disorder strength $V_{d}$. For
each fixed value of $V_{z}$, we varied $V_{d}$ to infer whether the soliton
would survive after 10 periods of the oscillations. The goal was to compute
the \textit{survival rate} of the soliton under the influence of the
disorder, using 10 realizations for this purpose. In each realization, the
survival of the soliton was registered if its final norm, integrated over
the FWHM\ range around its center, exceeded $50\%$ of the initial value in
the same range. The survival rate is then defined as the number of
realizations featuring the surviving soliton, divided by 10 (the total
number of the realizations). Typical examples of the survival and
destruction of the oscillating solitons are displayed in Fig. \ref{f:VD5_180}.
[Note that $h$ in the horizontal axis label denotes the Planck constant].

\begin{figure}[tbp]
\centerline{
    \psfig{file = 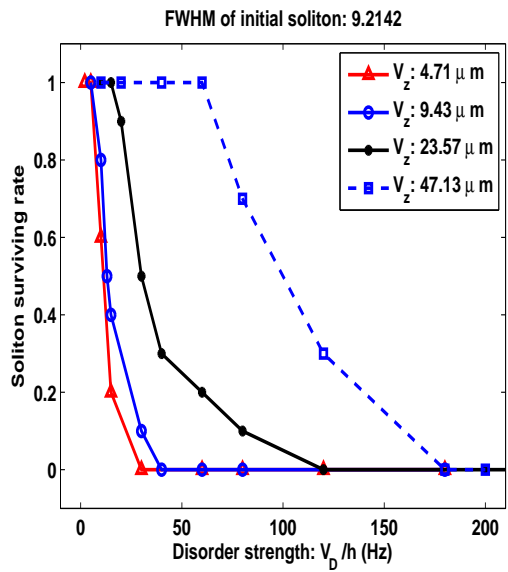, width=3in,height=3in}
      }
\caption{(Color online) The survival rate of the oscillating soliton for
initial shift $x_{0}=0.6$ mm.}
\label{f:X600rate}
\end{figure}

\begin{figure}[tbp]
\centerline{
    \psfig{file = 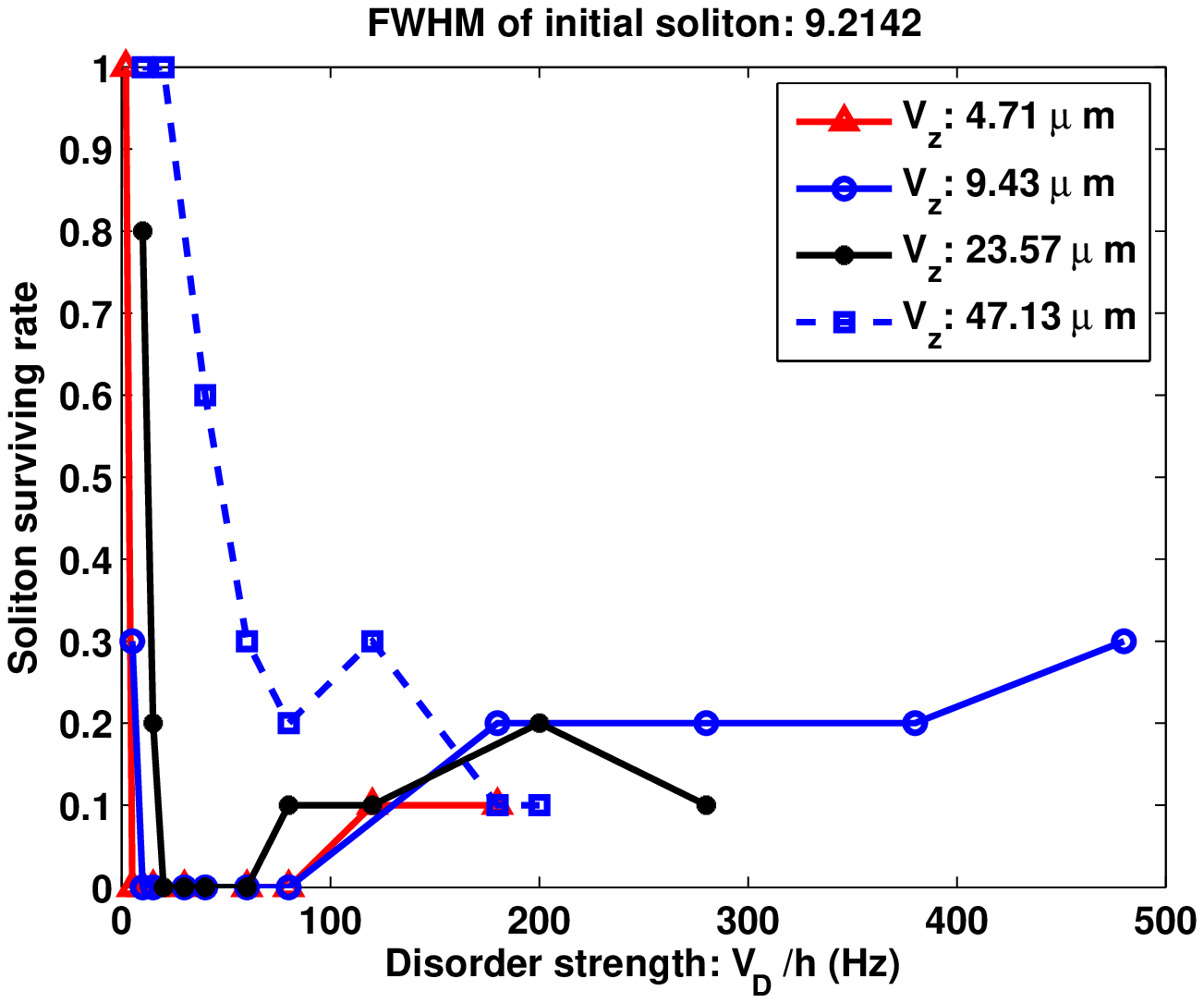, width=3in,height=3in}
      }
\caption{(Color online) The survival rate of the oscillating soliton for $%
x_{0}=0.1$ mm. }
\label{f:X100rate}
\end{figure}

The results for $x_{0}=0.6$ and $0.1$ mm are displayed in Figs.~%
\ref{f:X600rate} and \ref{f:X100rate}. They correspond to the dimensionless values 
of $x_{0}=254.6$ and $42.4$ in Eq.(\ref{NLS}).
 These clearly indicate that the survival rate drops to zero as the disorder
strengthens and the correlation length decreases, in agreement with the
experimental observations for the repulsive case
\cite{Randy3}. This is a natural consequence of the
increasing rate of the emission of radiation by the soliton oscillating
across the random potential.

\begin{figure}[tbp]
\centerline{
    \psfig{file = 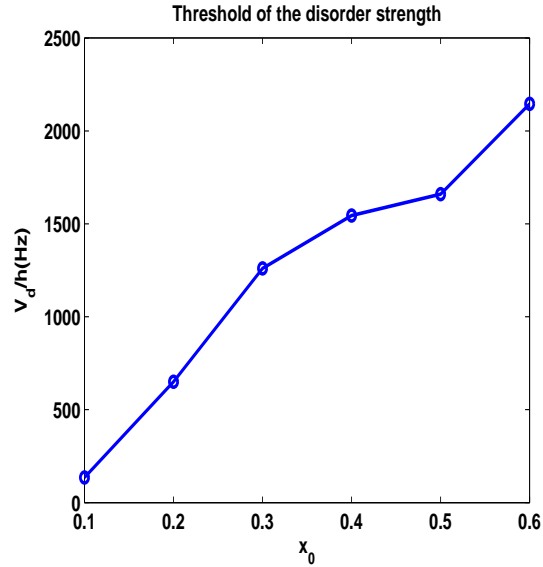, width=3in,height=3in}
      }
\caption{(Color online) For given initial shift $x_{0}$, the soliton
survives if the disorder strength, $V_{d}$, exceeds the minimum value shown
in this plot.}
\label{f:VD_vs_x0}
\end{figure}

When the disorder becomes very strong, the soliton starts to survive again,
as seen in Fig.~\ref{f:X100rate} for $x_{0}=0.1$ mm. This is explained by
the fact that a very strong random potential consists of local potential
wells which quickly trap and immobilize the soliton, preventing its decay
into radiation and, in addition, the strong random potential impedes the
separation of the radiation waves from the parent soliton. The
restabilization boundary is shown in Fig.~\ref{f:VD_vs_x0}. The
approximately linear dependence of the minimum disorder strength, necessary
for the restabilization, on $x_{0}$ can be explained with the help the above
argument: the driving force acting on the soliton in the HO potential grows
linearly with $x_{0}$, while the largest pinning force, induced by the
random potential, is proportional to $V_{d}$. Therefore, the equilibrium
between the two, which determines the restabilization threshold, implies $%
V_{d}\sim x_{0}$.

\section{Conclusions}

The objective of this work was to report results of a systematic numerical
analysis of individual soliton and soliton train
dynamics in the experimentally relevant setting
based on the 1D NLSE including the combination of the HO
(harmonic-oscillator) and random potentials. This setting may be realized
in BEC and nonlinear optics. Two basic problems were considered: the
generation of soliton trains from the initial quasi-uniform state by the MI
(modulational instability), after the sudden switch of the nonlinearity from
the self-repulsion into attraction, and the survival or decay of a single
soliton oscillating in this combined potential. Dependences of the main
characteristics of the soliton train, and of the survival rate of the single
oscillating soliton, on the strength and correlation length of the
disordered potential, and also dependences of the characteristics of the
soliton array on the total norm of the initial state, have been produced by
averaging over a large number of random realizations. Salient features of
this dependences have been explained in a qualitative form.

A challenging problem is extending the analysis to the 2D setting, which may
suggest new possibilities for the experiment. There, one has to
consider
the interplay of the above analyzed mechanisms with the potential
collapse type events of super-critical atomic blobs and hence the
relevant phenomenology will be considerable richer. The 2D setting
is also of interest for the repulsive interaction dynamical case,
whereby the formation of dark solitons and trains thereof reported
in~\cite{Randy3} may be substituted by the formation of vortices
and vortex streets. 

{\bf Acknowledgment:} The first author's work was supported in part by
the National Science Foundation under the grant DMS-1016047. The second author
was supported by the National Science Foundation under the grant
DMS-0806762 and by the Alexander von Humboldt Foundation.


\begin{thebibliography}{99}
\bibitem{review} Kivshar Y S and Malomed B A, \textit{Dynamics of solitons
in nearly integrable systems}, 1989, \textit{Rev. Mod. Phys}. \textbf{61}
763.

\bibitem{review0} Gredeskul S A and Kivshar Y S, \textit{Propagation and
scattering of nonlinear waves in disordered systems}, 1992, \textit{Phys. Rep%
}. \textbf{216} 1-61.

\bibitem{review1} Belitz D and Kirkpatrick T R, \textit{The Anderson-Mott
transition}, 1994, \textit{Rev. Mod. Phys}. \textbf{66} 261-390.

\bibitem{review2} Hennig D and Tsironis G P, \textit{Wave transmission in
nonlinear lattices}, 1999, \textit{Phys. Rep}. \textbf{307}, 333-432.

\bibitem{review3} Evers F and Mirlin A D, \textit{Anderson transitions},
2008, \textit{Rev. Mod. Phys}.\textbf{\ 80} 1355-1417.

\bibitem{soliton1} Scharf R and Bishop A R, \textit{Length-scale competition
for the one-dimensional nonlinear Schr\"{o}dinger equation with spatially
periodic potentials}, 1993, Phys. Rev. B \textbf{47} 1375-1383.

\bibitem{soliton2} Alexeeva N V, Barashenkov I V, and Tsironis G P,
Impurity-induced stabilization of solitons in arrays of parametrically
driven nonlinear oscillators, 2000, Phys. Rev. Lett. \textbf{84} 3053-3056.

\bibitem{soliton3} Pertsch T, Peschel U, Kobelke, J Schuster K, Bartelt H,
Nolte S, T\"{u}nnermann A, and Lederer F, \textit{Nonlinearity and disorder
in fiber arrays}, 2004, Phys. Rev. Lett. \textbf{93} 053901.

\bibitem{soliton4} Schwartz T, Bartal G, Fishman S, and Segev M, \textit{%
Transport and Anderson localization in disordered two-dimensional photonic
lattices}, 2007 \textit{Nature} \textbf{446} 52-55.

\bibitem{review4} Kartashov Y V, Malomed B\ A and Torner L, \textit{Solitons
in nonlinear lattices}, 2011, \textit{Rev. Mod. Phys}. \textbf{83} 247-305.

\bibitem{BEC0} Wang D W, Lukin M D, and Demler E, \textit{Disordered
Bose-Einstein condensates in quasi-one-dimensional magnetic microtraps},
2004, \textit{Phys. Rev. Lett}. \textbf{92} 076802.

\bibitem{BEC1} Paul T, Leboeuf P, Pavloff N, Richter K, and Schlagheck P,
\textit{Nonlinear transport of Bose-Einstein condensates through waveguides
with disorder}, 2005, \textit{Phys. Rev. A} \textbf{72} 063621.

\bibitem{BEC2} Schulte T, Drenkelforth S, Kruse J, Ertmer W, Arlt J, Sacha
K, Zakrzewski J, and Lewenstein M, \textit{Routes towards Anderson-like
localization of Bose-Einstein condensates in disordered optical lattices},
2005, \textit{Phys. Rev. Lett}. \textbf{95} 170411.

\bibitem{BEC3} Shapiro B, \textit{Expansion of a Bose-Einstein condensate in
the presence of disorder}, 2007, \textit{Phys. Rev. Lett}. \textbf{99}
060602.

\bibitem{BEC4} Akkermans E, Ghosh S, and Musslimani Z H, \textit{Numerical
study of one-dimensional and interacting Bose-Einstein condensates in a
random potential}, 2008, \textit{J. Phys. B: At. Mol. Opt. Phys}. \textbf{41}
045302.

\bibitem{BEC5} Albert M, Paul T, Pavloff N and Leboeuf P, \textit{Dipole
oscillations of a Bose-Einstein condensate in the presence of defects and
disorder}, 2008, \textit{Phys. Rev. Lett}. \textbf{100} 250405.

\bibitem{BEC6} Fishman S, Krivolapov Y and Soffer A, \textit{On the problem
of dynamical localization in the nonlinear Schr\"{o}dinger equation with a
random potential}, 2008, \textit{J. Stat. Phys}. \textbf{131} 843-865.

\bibitem{Randy1} Chen, Y P, Hitchcock J, Dries D, Junker M, Welford C,
Pollack S E, Corcovilos T A, and Hulet R G, \textit{Phase coherence and
superfluid-insulator transition in a disordered Bose-Einstein condensate},
2008, \textit{Phys. Rev. A} \textbf{77} 033632.

\bibitem{Randy2} \ Chen, Y P, Hitchcock J, Dries D, Junker M, Welford C,
Pollack S E, Corcovilos T A, and Hulet R G, \textit{Experimental studies of
Bose-Einstein condensates in disorder}, 2009, \textit{Physica D} \textbf{238}
1321-1325.

\bibitem{Randy3} Dries D, Pollack S E, Hitchcock J M and Hulet R G, \textit{%
Dissipative transport of a Bose-Einstein condensate}, 2010, \emph{Phys. Rev.
A} \textbf{82} 033603.

\bibitem{GPE} Pitaevskii L P and Stringari A 2003 \textit{Bose-Einstein
Condensation }(Clarendon Press, Oxford).

\bibitem{Montesinos05} Montesinos G D and P\'{e}rez-Garc\'{\i}a V M, \textit{%
Numerical studies of stabilized Townes solitons}, 2005 \textit{Math. Comput.
Simul}. \textbf{69} 447-456.

\bibitem{Blanes02} Blanes S and Moan P C, \textit{Practical symplectic
partitioned Runge-Kutta and Runge-Kutta-Nystrom methods}, 2002, \textit{J.
Comp. Appl}. \textit{Math}. \textbf{142} 313-330.

\bibitem{Yaglom} Yaglom A M
2004  \textit{Introduction to the Theory of Stationary Random Functions}
(Dover Phoenix Editions ) .

\bibitem{imaginary} Chiofalo M L, Succi S, and Tosi M P, \textit{Ground
state of trapped interacting Bose-Einstein condensates by an explicit
imaginary-time algorithm}, 2000, \textit{Phys. Rev. E} \textbf{62} 7438-7444.

\bibitem{Randy} Strecker K E, Partridge G B, Truscott A G and Hulet R G,
\textit{Formation and propagation of matter-wave soliton trains}, 2002,
\textit{Nature} \textbf{417} 150-153.

\bibitem{Zakharov} Zakharov V E, Manakov S\ V, Novikov S\ P and Pitaevskii L
P 1984 \textit{Theory of Solitons: the Inverse Scattering Method}
(Consultants Bureau, New York) .

%\bibitem{Yaglom} Yaglom A M
%2004  \textit{Introduction to the Theory of Stationary Random Functions}
%(Dover Phoenix Editions ) .

%\bibitem{Zakharov} V. E. Zakharov, S. V. Manakov, S. P. Novikov, and L. P.
%Pitaevskii, \textit{Solitons: Inverse Scattering Transform}, Nauka
%Publishers (Moscow, 1980) [English translation: Consultants Bureau, New
%York, 1984].
%
%\bibitem{ablowitz} M. J. Ablowitz and H. Segur, \textit{Solitons and the
%Inverse Scattering Transform}, SIAM (Philadelphia, 1981); M. J. Ablowitz and
%P. A. Clarkson, \textit{Solitons, Nonlinear Evolution Equations and Inverse
%Scattering}, Cambridge University Press (Cambridge, 1991).
%
%\bibitem{ablowitz1} M. J. Ablowitz, B. Prinari and A. D. Trubatch, \textit{%
%Discrete and Continuous Nonlinear Schr{\"{o}}dinger Systems}, Cambridge
%University Press (Cambridge, 2004).
%
%\bibitem{sulem} C. Sulem and P. L. Sulem,
%\newblock  {\it The Nonlinear
%Schr{\"o}dinger Equation}, Springer-Verlag (New York, 1999).
%
%\bibitem{hasegawa} A. Hasegawa, \textit{Solitons in Optical Communications},
%Clarendon Press (Oxford, NY 1995).
%
%\bibitem{malomed} B. A. Malomed,
%%{\it Variational methods in nonlinear fiber optics and related fields},
%Progress in Optics \textbf{43}, (2002) 69-191.
%
%\bibitem{zakh} V. E. Zakharov, %{\it Collapse of Langmuir waves},
%Sov. Phys. JETP \textbf{35} (1972) 908-914; V. E. Zakharov,
%%{\it Collapse and Self-focusing of Langmuir Waves},
%\newblock Handbook of Plasma Physics, (M. N. Rosenbluth and R. Z. Sagdeev,
%eds.), vol. 2 (A. A. Galeev and R. N. Sudan eds.), pp. 81-121, Elsevier
%(1984).
%
%\bibitem{benjamin} T. B. Benjamin and J. E. Feir,
%%{\it The disintegration of wavetrains in deep
%%water, Part 1},
%J. Fluid Mech. \textbf{27}, (1967) 417-430; M. Onorato, A. R. Osborne, M.
%Serio, and S. Bertone %{\it Freak waves in random oceanic sea states},
%Phys. Rev. Lett. \textbf{86} (2001) 5831.
%
%\bibitem{magnet} J. Corones, Phys. Rev. B \textbf{16}, 1763 (1977); B. A.
%Ivanov, A. M. Kosevich, and I. V. Manzhos, Solid State Commun. \textbf{34},
%417 (1980); A. M. Kosevich, B. A. Ivanov, and A. S. Kovalev, Phys. Rep.
%\textbf{194}, 117 (1990).
%
%\bibitem{yannac} G. P. Flessas, P. G. L. Leach and A. N. Yannacopoulos, J.
%Opt. B \textbf{6}, S161 (2004).
%
%\bibitem{abdullaev} F. Kh. Abdullaev, \textit{Theory of solitons in
%inhomogeneous media}, John Wiley and Sons (New York, 1994).
%
%\bibitem{pethick} C. J. Pethick and H. Smith, \textit{Bose-Einstein
%condensation in dilute gases}, Cambridge University Press (Cambridge, 2002);
%L. P. Pitaevskii and S. Stringari, \textit{Bose-Einstein Condensation},
%Oxford University Press (Oxford, 2003).
%
%\bibitem{Abd} F. Kh. Abdullaev, J. C. Bronski, and R. M. Galimzyanov,
%Physica D \textbf{184}, 319 (2003).
%
%\bibitem{CQ} F. Kh. Abdullaev, J. Garnier, Phys. Rev. E \textbf{72} (2005)
%035603 (R).
%
%\bibitem{inguscio} J. E. Lye, L. Fallani, M. Modugno, D. Wiersma, C. Fort
%and M. Inguscio, Phys. Rev. Lett. \textbf{95}, 070401 (2005).
%%cond-mat/0412167.
%; C. Fort, L. Fallani, V. Guarrera, J. Lye, M. Modugno, D. S. Wiersma and M.
%Inguscio, Phys. Rev. Lett. \textbf{95}, 170410 (2005); L. Fallani, J. E.
%Lye, V. Guarrera, C. Fort and M. Inguscio, Phys. Rev. Lett. \textbf{98},
%130404 (2007).
%
%\bibitem{Randy} Y. P. Chen, J. Hitchcock, D. Dries, M. Junker, C. Welford,
%and R. G. Hulet,
%%Phase coherence and superfluid-insulator transition in a disordered Bose-Einstein condensate
%Phys. Rev. A \textbf{77}, 033632 (2008).
%
%\bibitem{Abd2} F. Kh. Abdullaev and J. Garnier,
%%Propagation of matter-wave solitons in periodic and random nonlinear potentials
%Phys. Rev. A \textbf{72}, 061605 (2005).
%
%\bibitem{ziad} E. Akkermans, S. Ghosh and Z. H. Musslimani, J. Phys. B
%\textbf{41}, 045302 (2008).
%
%\bibitem{Moti} T. Schwartz, G. Bartal, S. Fishman, and M. Segev,
%%Transport and Anderson localization in disordered two-dimensional photonic lattices,
%Nature \textbf{446} , 52-55 (2007); Y. Lahini, A. Avidan, F. Pozzi, M.
%Sorel, R. Morandotti, D. N. Christodoulides, and Y. Silberberg,
%%Anderson localization and nonlinearity in one-dimensional disordered photonic lattices,
%Phys. Rev. Lett. \textbf{100}, 013906 (2008).
%
%\bibitem{Koehler} T. K\"ohler, K. Goral and P. S. Julienne, Rev. Mod. Phys.%
%\textbf{78}, 1311 (2006).
%
%\bibitem{feshbachNa} S. Inouye, M. R. Andrews, J. Stenger, H. J. Miesner, D.
%M. Stamper-Kurn and W. Ketterle, Nature \textbf{392}, 151 (1998); J.
%Stenger, S. Inouye, M. R. Andrews, H.-J. Miesner, D. M. Stamper-Kurn, and W.
%Ketterle, Phys. Rev. Lett. \textbf{82}, 2422 (1999); J. L. Roberts, N. R.
%Claussen, J. P. Burke Jr., C. H. Greene, E. A. Cornell, and C.~E. Wieman,
%Phys. Rev. Lett. \textbf{81}, 5109 (1998); S.~L. Cornish, N. R. Claussen, J.
%L. Roberts, E. A. Cornell, and C. E. Wieman, Phys. Rev. Lett. \textbf{85},
%1795 (2000).
%
%\bibitem{ofr} F. K. Fatemi, K. M. Jones, and P. D. Lett, Phys. Rev. Lett.
%\textbf{85}, 4462 (2000); M. Theis, G. Thalhammer, K. Winkler, M. Hellwig,
%G. Ruff, R. Grimm, and J. H. Denschlag, Phys. Rev. Lett. \textbf{93}, 123001
%(2004).
%
%\bibitem{book} B. A. Malomed, \textit{Soliton Management in Periodic Systems}%
%, Springer (New York, 2006).
%
%\bibitem{expb1} K.~E.\ Strecker, %\textit{et al.},
%G. B.\ Partridge, A. G.\ Truscott, and R. G.\ Hulet, Nature \textbf{417},
%150 (2002).
%
%\bibitem{expb2} L.\ Khaykovich, %\textit{et al.},
%F.\ Schreck, G.\ Ferrari, T.\ Bourdel, J.\ Cubizolles, L. D.\ Carr, Y.\
%Castin, and C.\ Salomon, Science \textbf{296}, 1290 (2002).
%
%\bibitem{expb3} S. L. Cornish, S. T. Thompson, and C. E. Wieman, Phys.\
%Rev.\ Lett.\ \textbf{96}, 170401 (2006).
%
%\bibitem{molecule} J. Herbig, T. Kraemer, M. Mark, T. Weber, C. Chin, H. C.
%Nagerl, and R. Grimm, Science \textbf{301}, 1510 (2003); C.~A. Regal, C.
%Ticknor, J. L. Bohn, and D. S. Jin, Nature \textbf{424}, 47 (2003).
%
%\bibitem{becbcs} M. Bartenstein, A. Altmeyer, S. Riedl, S. Jochim, C. Chin,
%J. H. Denschlag, and R. Grimm, Phys. Rev. Lett. \textbf{92}, 203201 (2004);
%T. Bourdel, L. Khaykovich, J. Cubizolles, J. Zhang, F. Chevy, M. Teichmann,
%L. Tarruell, S. J. J. M. F. Kokkelmans, and C. Salomon, Phys. Rev. Lett.
%\textbf{93}, 050401 (2004).
%
%\bibitem{FRM1} F.~Kh. Abdullaev, J. G. Caputo, R. A. Kraenkel, and B. A.
%Malomed, Phys. Rev. A \textbf{67}, 013605 (2003); H. Saito and M. Ueda,
%Phys. Rev. Lett. \textbf{90}, 040403 (2003); G.~D. Montesinos, V. M.
%P\'erez-Garc\'{\i}a, and P. J. Torres, Physica D \textbf{191} 193 (2004).
%
%\bibitem{FRM2} P.~G. Kevrekidis, G. Theocharis, D. J. Frantzeskakis, and B.
%A. Malomed, Phys. Rev. Lett. \textbf{90}, 230401 (2003); D.~E. Pelinovsky,
%P. G. Kevrekidis, and D. J. Frantzeskakis, Phys. Rev. Lett. \textbf{91},
%240201 (2003); D.~E. Pelinovsky, P. G. Kevrekidis, D. J. Frantzeskakis, and
%V. Zharnitsky, Phys. Rev. E \textbf{70}, 047604 (2004); Z.~X. Liang, Z. D.
%Zhang, and W. M. Liu, Phys. Rev. Lett. \textbf{94}, 050402 (2005); M.
%Matuszewski, E. Infeld, B. A. Malomed, and M. Trippenbach, Phys. Rev. Lett.
%\textbf{95}, 050403 (2005).
%
%\bibitem{Warsaw} M. Trippenbach, M. Matuszewski, and B. A. Malomed,
%%Stabilization of three-dimensional matter-waves solitons in an optical lattice
%Europhys. Lett. \textbf{70}, 8 (2005); M. Matuszewski, E. Infeld, B. A.
%Malomed, and M. Trippenbach, Phys. Rev. Lett. \textbf{95}, 050403 (2005).
%
%\bibitem{Isaac} I. Towers and B. A. Malomed, J. Opt. Soc. Am. B \textbf{19},
%537 (2002).
%
%\bibitem{karnia} G. Lin, L. Grinberg and G. E. Karniadakis, J. Comp. Phys.
%\textbf{213}, 676 (2006).
%
%\bibitem{Wiener38} N. Wiener, Amer. J. Math. \textbf{60}, 897 (1938).
%
%\bibitem{Loeve77} M. Lo\`{e}ve, \textit{Probability Theory}, Springer-Verlag
%(New York, 1977).
%
%\bibitem{XiuDB03} D. Xiu and G. E. Karniadakis, J. Comp. Phys. \textbf{187},
%137 (2003).
%
%\bibitem{Ghanem91} R. G. Ghanem and P. D. Spanos, \textit{Stochastic Finite
%Element: A Spectral Approach}, Springer-Verlag (New York, 1991).
%
%\bibitem{townes} R. Y. Chiao, E. Garmire, and C. H. Townes Phys. Rev. Lett.
%\textbf{13}, 479 (1964).
%
%\bibitem{Itin} A. Itin, T. Morishita, and S. Watanabe,
%%Reexamination of dynamical stabilization of matter-wave solitons
%Phys. Rev. A \textbf{74}, 033613 (2006).
%
%%\bibitem{XiuDB03} D. Xiu and G. E. Karniadakis, J. Comp. Phys. \textbf{187},
%%137 (2003).
%
%%\bibitem{Ghanem91} R. G. Ghanem and P. D. Spanos, \textit{Stochastic Finite
%%Element: A Spectral Approach}, Springer-Verlag (New York, 1991).
%
%\bibitem{CQstability} D. E. Pelinovsky, Y. S. Kivshar, V. V. Afanasjev,
%Physica D \textbf{116} (1998) 121; L. Khaykovich,d B. A. Malomed, Phys. Rev.
%A \textbf{74} (2006) 023607.
%
%\bibitem{malomed1} B. B. Baizakov, B.A. Malomed and M. Salerno, Europhys.
%Lett. \textbf{63}, 642 (2003); J. Yang and Z. H. Musslimani, Opt. Lett.
%\textbf{28}, 2094 (2003); B. B. Baizakov, B. A. Malomed and M. Salerno,
%Phys. Rev. A \textbf{70}, 053613 (2004).
\end{thebibliography}
\end{document}